\documentclass[12pt,english,floatfix,nofootinbib,superscriptaddress,aps,prd,preprint]{revtex4}
\usepackage[utf8]{inputenc}
\usepackage{float}
\usepackage{array}
\usepackage{lipsum}
\usepackage{graphicx}
\usepackage{amsmath,amsthm,amsfonts,amssymb}
\usepackage{tikz}
\usepackage{multirow}
\usepackage{appendix}
 \usetikzlibrary{quotes,angles}
 \usetikzlibrary{arrows} 
\usetikzlibrary{decorations.markings}
\usepackage{graphicx}
\usepackage[english]{babel}
\usepackage{color}
\usepackage{subfig}
\usepackage{caption}
\usepackage{tensor}
\usepackage{esint}
\usepackage[dvips]{epsfig}
\usepackage[dvips]{graphicx}
\usepackage{float}
\usepackage{units}
\usepackage{textcomp}
\usepackage{mathrsfs}
\usepackage{amsmath}
\usepackage[makeroom]{cancel}
\usepackage{amssymb}
\usepackage{amsbsy}
\usepackage{amsfonts}
\usepackage{amssymb,mathrsfs,xcolor}
\usepackage{esint}
\usepackage{array}
\usepackage{graphicx}

\usepackage{hyperref}
\hypersetup{
    colorlinks,
    citecolor=blue,
    filecolor=green,
    linkcolor=purple,
    urlcolor=red,
}

\usepackage{slashed}

\newcommand{\ie}{\begin{equation}}
\newcommand{\fe}{\end{equation}}
\newcommand{\se}{\begin{eqnarray}}
\newcommand{\ff}{\end{eqnarray}}

\usepackage{hyperref}
\hypersetup{colorlinks,breaklinks,
			citecolor=[rgb]{0,0.0,1.0},
            urlcolor=[rgb]{0.0,0.0,1.0},
            linkcolor=[rgb]{0,0.5,0.9}}

\begin{document}

\title{Gravitational lensing by a Lorentz-violating black hole}

\author{A. A. Ara\'{u}jo Filho}
\email{dilto@fisica.ufc.br}
\affiliation{Departamento de Física, Universidade Federal da Paraíba, Caixa Postal 5008, 58051-970, João Pessoa, Paraíba,  Brazil.}
\affiliation{Departamento de Física, Universidade Federal de Campina Grande Caixa Postal 10071, 58429-900 Campina Grande, Paraíba, Brazil}

\author{J. R. Nascimento}
\email{jroberto@fisica.ufpb.br}
\affiliation{Departamento de Física, Universidade Federal da Paraíba, Caixa Postal 5008, 58051-970, João Pessoa, Paraíba,  Brazil.}

\author{A. Yu. Petrov}
\email{petrov@fisica.ufpb.br}
\affiliation{Departamento de Física, Universidade Federal da Paraíba, Caixa Postal 5008, 58051-970, João Pessoa, Paraíba,  Brazil.}

\author{P. J. Porfírio}
\email{pporfirio@fisica.ufpb.br}
\affiliation{Departamento de Física, Universidade Federal da Paraíba, Caixa Postal 5008, 58051-970, João Pessoa, Paraíba,  Brazil.}




\date{\today}

\begin{abstract}

In this work, we study the gravitational lensing by a Lorentz-violating (LV) black hole in metric-affine bumblebee gravity in the framework of the Standard Model Extension, inspired by the recent contribution \cite{Filho:2022yrk}. Explicitly, we concentrate on a specific application: we perform the computation of gravitational lensing effects under the strong field limit. In particular, we analytically derive the deflection angle so that the lens equation can also be addressed. This methodological approach yields physically measurable outcomes, including the determination of relativistic image positions and their corresponding magnifications. As an application of this methodology, we consider the gravitational lensing by Sagittarius A${}^*$ and obtain the corresponding observables expressed as functions of the LV parameter.

\end{abstract}


\maketitle


\section{Introduction}
\label{sec:intro}

The bending of light as it moves through a gravitational field was one of the first predictions of general relativity (GR), which played a crucial role in proving its validity \cite{01,02}. Since then, gravitational lenses have become essential tools in astrophysics and cosmology \cite{03,04}, helping to study various aspects such as matter distributions \cite{05,06}, dynamics of dark matter \cite{07}, and many other topics \cite{08,09,010,011,012,013,014,015}.

Einstein's prediction, initially formulated within the weak field approximation, mainly deals with situations where light travels long distances from the source of a gravitational lens. Recent advancements, such as the detection of gravitational waves by the LIGO-Virgo collaboration \cite{016,017,018}, have opened up new aspects for exploration. This involves using gravitational waves to investigate cosmological aspects, including the study of gravitational lensing within the framework of the weak field approximation \cite{019,020}.

In regions with intense gravitational fields, where light rays passe very closely to the source of the lens, often represented by compact objects like a black hole, the bending of light is greatly magnified. The early studies on gravitational lenses within this strong field scenario can be traced back to significant works such as the Schwarzschild spacetime analysis \cite{021}, which was later expanded to include general spherically symmetric and static spacetimes \cite{022}.

In recent years, there has been a significant increase in researches in this field, particularly driven by experimental findings indicating the presence of a supermassive black hole at the center of our Milky Way galaxy \cite{023}. Notably, the groundbreaking observations made by the Event Horizon Telescope (EHT) collaboration, revealing the silhouette of a black hole at the core of the galaxy M87, have spurred these efforts \cite{024,025,026,029,EventHorizonTelescope:2019dse}. As a further continuation of these studies, observations of the Sagittarius A${}^*$ (Sgr A${}^*$) \cite{EventHorizonTelescope:2022wkp,EventHorizonTelescope:2022xqj} certainly worth mentioning. As a result, due to the extremely powerful gravitational fields around such objects, relying on the weak field approximation is no longer feasible, necessitating a robust approach suited to strong field conditions.

The mathematical treatment of gravitational lenses in these scenarios poses significant challenges. One of the first who tackled this issue was Synge \cite{Synge:1966okc}, who investigated the escape of photons from stars with strong gravitational fields. Indeed, he derived an equation for the minimal angle at which photons are not recaptured by the compact object. This equation can also be used to calculate the angular radius of the shadow (the region inside the photon ring) of the Schwarzschild black hole \cite{Luminet:1979nyg}. Recent collective efforts have made more progress towards developing comprehensive analytical frameworks. For instance, Virbhadra and Ellis formulated a concise lens equation for supermassive black holes in galaxies, considering an asymptotically flat background. This equation effectively addressed substantial light deflections \cite{030,Virbhadra:2022iiy, Virbhadra:2008ws, 031}. Their findings revealed a multitude of symmetrically distributed images around the optic axis, a result of the strong gravitational influence on light rays. Additionally, further studies examining images produced by lensing from black holes, notably the Kerr black hole, are also available \cite{Aratore:2024ilt}.

Additional research has expanded our analytical methods. Fritelli and others \cite{032} developed a precise lens equation for the Schwarzschild spacetime without specifying a background. Bozza et al. \cite{bozza2001strong} independently derived an expression for light deflection in strong fields, particularly near the photon sphere, and later applied it to the Schwarzschild spacetime \cite{034}. Subsequent refinements by Tsukamoto \cite{tsukamoto2017deflection} have built upon these remarks. The study of light deviation in strong fields has expanded to various contexts, including the Reissner-Nordström spacetime \cite{036,036.1,036.2}, rotating solutions \cite{37.1,37.2,37.3,37.4,37.5,37.6,Hsieh:2021scb,Hsieh:2021rru}, exotic constructs like wormholes \cite{38.1,38.2,38.3,38.4,38.5}, and topological defects \cite{39}, as well as investigations within modified gravity theories \cite{40}, including the scalar-tensor-vector gravity \cite{Moffat:2014aja,Moffat:2015kva}, and studies of regular black holes \cite{41}, naked singularities \cite{42}, and braneworld scenarios \cite{43}. In this context, it is worth calling attention to the seminal paper \cite{Cavalcanti:2016mbe}, where the authors have investigated the gravitational lensing effects in the strong field regime for Casadio-Fabbri-Mazzacurati braneworld black hole metrics \cite{Casadio:2001jg}. Additionally, other studies have explored gravitational lensing in various spacetimes. For instance, some studies explored light bending in black--bounce spacetimes \cite{Nascimento:2020ime, Ghosh:2022mka, AbhishekChowdhuri:2023ekr}, while others focused on spacetimes with topological charge \cite{Furtado:2020puz,Soares:2023err}. Another investigation involved a holonomy--corrected Schwarzschild spacetime \cite{Soares:2023uup}, and a separate study examined Schwarzschild--like black holes within metric-affine bumblebee gravity \cite{Lambiase:2023zeo}.

The Standard Model Extension (SME), when extended to include gravitational effects, provides a comprehensive framework that incorporates terms allowing for the violation of Lorentz/CPT symmetry \cite{kostelecky2004gravity}. For a general overview of Lorentz/CPT violation coefficients in modified gravity, we indicate \cite{Kostelecky:2020hbb}. Within the gravitational domain, the SME operates within a Riemann--Cartan manifold, where torsion is treated as a dynamic geometrical quantity alongside the metric. While it is possible to introduce non--Riemannian terms within the gravitational SME sector, previous studies have mainly focused on the metric approach to gravity, where the metric serves as the sole dynamical geometric field.

The metric-affine (Palatini) formalism presents an extension formalism of the metric approach, treating the metric and connection as independent dynamic geometric entities \cite{Ghil1, Ghil2}. Recent contributions have been considered recently in within the context of bumblebee gravity scenarios \cite{Paulo2, Paulo3, Paulo4}. In particular, an exact Schwarzschild--like solution has been proposed in the literature \cite{Filho:2022yrk}, accompanied by estimates for the Lorentz violation (LV) parameter derived from classical gravitational tests. Additionally, Refs. \cite{Lambiase:2023zeo, Jha:2023vhn, hassanabadi2023gravitational} has investigated the shadow and quasinormal modes of this black hole, offering a comprehensive examination of its astrophysical implications. However, there is currently a gap in the literature regarding gravitational lensing via the strong field approximation. Therefore, our focus is devoted to a specific metric-affine bumblebee gravity model. This model is linked to the LV coefficients of the Standard Model Extension assuming nontrivial values for $u$ and $s^{\mu\nu}$, while $t^{\mu\nu\alpha\beta}=0$. (This absence of the $t$ coefficient, often referred to as the $t$-puzzle, has been discussed in \cite{Bonder:2015maa}).

The paper unfolds as follows: in Sec. \ref{Application}, we delve into a comprehensive overview of the strong field limit, accompanied by novel computations for the vacuum solution of bumblebee gravity within the metric--affine formalism \cite{Filho:2022yrk}, in Sec. III we study the deflection angle within the gravitational lensing produced by the LV black hole whose metric has been obtained in our previous paper \cite{Filho:2022yrk}, and in Sec. IV we explicitly evaluate some parameters associated with the light deviation. Concluding our discourse, Sec. \ref{summary} encapsulates our key findings and draws overarching conclusions.


\section{Gravitational lensing by a static and  spherically symmetric metric}
\label{Application}

 In this section, our interest is to provide a mathematical expression for the deflection angle suffered by a photon (light), originating from the infinity $(r\rightarrow \infty)$, as approaches a gravitational object. Such a phenomenon is called gravitational lensing. In this work, it is worth remarking that we will focus the study of gravitational lensing on the strong field regime, which is achieved when the photon skims very closely to the source of the lens (massive object). Yet, different from the weak field regime, an involved mathematical treatment is needed, which will be briefly described below.

  We shall use the methodology employed in \cite{tsukamoto2017deflection} to study the deflection angle in the strong field regime. This method is only valid for a class of asymptotically flat, static, and spherically symmetric spacetimes which are characterized by the line element:  

\ie
\label{ssst}
\mathrm{d}s^{2} = - A(r) \mathrm{d}t^{2} + B(r) \mathrm{d}r^{2} + C(r)(\mathrm{d}\theta^2 + \sin^{2}\theta\mathrm{d}\phi^2),
\fe
where $\lim\limits_{r \to \infty}A(r) = 1$, $\lim\limits_{r \to \infty}B(r) = 1$, and $\lim\limits_{r \to \infty}C(r) = r^{2}$. The spacetime symmetries lead to two Killing vectors, namely, $\partial_t$ and $\partial_\phi$, and, consequently, to the existence of two conserved quantities: the energy $E$ and the angular momentum $L$ (see \cite{Carroll:2004st} for more details). 

Following this methodology, it is important to define some quantities, namely, the closest distance that the photon passes through the gravitational object, denoted as $r_0$, the radius of the photon sphere (that is, the radius of the stable of a photon), denoted as $r_m$. The strong field regime is attained when $r_0\rightarrow r_m$. It is also useful to define the impact parameter
\ie
b \equiv \frac{L}{E} = \frac{C(r)\Dot{\phi}}{A(r)\Dot{t}},
\fe

Given the symmetry properties of the aforementioned metric, the geodesic trajectories of photons (see \cite{tsukamoto2017deflection} for a detailed discussion) reduce to
\ie
\left(  \frac{\mathrm{d}r}{\mathrm{d}\phi}     \right)^{2} = \frac{R(r)C(r)}{B(r)},
\fe
where $R(r) \equiv C(r)/A(r) b^{2} -1$ and the deflection angle of light, $\alpha(r_{0})$ reads
\ie
\alpha(r_{0}) = I(r_{0}) - \pi.
\fe
Here, $I(r_{0})$ can be defined as 
\ie
I(r_{0}) \equiv 2 \int^{\infty}_{r_{0}} \frac{\mathrm{d}r}{\sqrt{\frac{R(r)C(r)}{B(r)}}}.
\fe
As proposed by Tsukamoto \cite{tsukamoto2017deflection}, it is convenient to define the new variable
\ie
z \equiv 1 - \frac{r_{0}}{r},
\fe
in a such way that we redefine the integral as
\ie
\label{int}I(r_{0}) = \int^{1}_{0} f(z,r_{0}) \mathrm{d}z,
\fe
with 
\ie
f(z,r_{0}) \equiv \frac{2r_{0}}{\sqrt{G(z,r_{0})}}, \,\,\,\,\,\,\,\, \text{and} \,\,\,\,\,\,\,\,  G(z,r_{0}) \equiv R \frac{C}{B}(1-z)^{4}.
\fe
The next step is to expand $G(z, r_0)$ in power series in $z$ and then take the limit $r_0\rightarrow r_m$ (the strong field regime). After some mathematical manipulation (see \cite{tsukamoto2017deflection}), we found that the integral $I(r_0)$ is logarithmically divergent (see for) and a regularization procedure is required. Roughly speaking, to deal with this divergence, we decompose the integral \(I(r_0)\) into two components: a divergent piece denoted as \(I_D(r_0)\) and a well-behaved one denoted as \(I_R(r_0)\). It is noteworthy that we are not concerned with the step-by-step calculation of $I_D(r_0)$, which can be found in \cite{tsukamoto2017deflection}. Then, the regular part of the integral \eqref{int} written in terms of the impact parameter reads
\ie
I_{R}(b) = \int^{1}_{0} f_{R}(z,b_{c})\mathrm{d}z + \mathcal{O}[(b-b_{c})\ln(b-b_{c})],
\fe
where \ie
b_{c}(r_{m}) \equiv \lim_{r_{0} \to r_{m}} \sqrt{\frac{C_{0}}{A_{0}}},
\fe
is the critical impact parameter, the subscript ``$m$'' denotes quantities evaluated at $r=r_0$. Moreover, $f_{R} \equiv f(z,r_{0}) - f_{D}(z,r_{0})$ and, in the strong field limit, the respective deflection angle is 
\ie
a(b) = - \Tilde{a} \ln \left[ \frac{b}{b_{c}}-1    \right] + \Tilde{b} + \mathcal{O}[(b-b_{c})\ln(b-b_{c})],
\label{deflections}
\fe
with
\ie
\Tilde{a} = \sqrt{\frac{2 B_{m}A_{m}}{C^{\prime\prime}_{m}A_{m} - C_{m}A^{\prime\prime}_{m}}}, \,\,\,\,\,\,\,\, \text{and} \,\,\,\,\,\,\,\, \Tilde{b} = \Tilde{a} \ln\left[ r^{2}_{m}\left( \frac{C^{\prime\prime}_{m}}{C_{m}}  -  \frac{A^{\prime\prime}_{m}}{C_{m}} \right)   \right] + I_{R}(r_{m}) - \pi.
\fe
In the next sections, we apply this methodology to the LV black hole metric obtained in \cite{Filho:2022yrk} and investigate the LV effects in the deflection angle.


\section{Application: gravitational lensing by a LV black hole}

The metric obtained in \cite{Filho:2022yrk} describes a non-rotating black hole which incorporates the effects of Lorentz symmetry breaking (LSB) (see the Appendix for some basic concepts of the LSB and the dependence of the results on different observer or particle frames). The metric was shown to look like  
\begin{equation}
    \mathrm{d}s^2_{(g)}=-\frac{\left(1-\frac{2M}{r}\right)}{\sqrt{\left(1+\frac{3X}{4}\right)\left(1-\frac{X}{4}\right)}}\mathrm{d}t^2+\frac{\mathrm{d}r^2}{\left(1-\frac{2M}{r}\right)}\sqrt{\frac{\left(1+\frac{3X}{4}\right)}{\left(1-\frac{X}{4}\right)^3}}+r^{2}\left(\mathrm{d}\theta^2 +\sin^{2}{\theta}\mathrm{d}\phi^2\right),
    \label{metric3}
\end{equation}
where $X$ stands for the LV coefficient. Earlier, some issues related to this black hole, including quasinormal modes and Hawking radiation, have been studied in \cite{Jha:2023vhn}. Now, by proceeding with the following redefinitions: 
\begin{equation}
\hat{t}\equiv \sqrt{\Delta_1}\,t,\,\, \hat{r}\equiv \sqrt{\Delta_2}\,r,\,\, \hat{M}\equiv \sqrt{\Delta_2}\, M,\,\, \hat{K}^2\equiv \dfrac{1}{\Delta_2}
\label{14}
\end{equation}
where
\begin{equation}
    \Delta_1 = \left[ \left( 1 + \frac{3X}{4}    \right) \left(  1 - \frac{X}{4}  \right) \right]^{-\frac{1}{2}}\,\,\mbox{and}\,\,\Delta_2 = \left[ \left( 1 + \frac{3X}{4}\right)  \left( 1 - \frac{X}{4}  \right)^{-3} \right]^{\frac{1}{2}},
    \label{15}
\end{equation}
the line element \eqref{metric3} can be set into a more compact form,
\begin{equation}
    \mathrm{d}s^2_{(g)}=-\left(1-\frac{2\hat{M}}{\hat{r}}\right)\mathrm{d}\hat{t}^{\,2}+\frac{1}{\left(1-\frac{2\hat{M}}{\hat{r}}\right)}\mathrm{d}\hat{r}^2+K^2\hat{r}^2(\mathrm{d}\theta^2+\sin^2(\theta)\mathrm{d}\phi^2).
    \label{metric4}
\end{equation}

The Eq. (\ref{metric4}) evidently replays the form (\ref{ssst}), with
\ie
A(\hat{r}) = \left(1-\frac{2\hat{M}}{\hat{r}}\right), \,\,\,\,\,\, B(\hat{r}) = \frac{1}{\left(1-\frac{2\hat{M}}{\hat{r}}\right)} \,\,\,\,\,\, \text{and} \,\,\,\,\,\, C(\hat{r}) = K^2 \hat{r}^{2}.
\fe
 It is noteworthy that in the asymptotic limit ($\hat{r}\rightarrow \infty$) the metric functions behave as: $\lim\limits_{\hat{r} \to \infty}A(\hat{r}) = 1$, $\lim\limits_{\hat{r} \to \infty}B(\hat{r}) = 1$, and $\lim\limits_{\hat{r} \to \infty}C(\hat{r}) = K^2\hat{r}^{2}$. At first sight, it is in contrast to the asymptotic flatness condition imposed in the previous section. However, taking into account that $X$ must be small since it represents the LV coefficient and $K^2$ is a positively definite constant, the remaining metric, in this limit, reduces to

\begin{equation}
  \lim\limits_{\hat{r} \to \infty}\mathrm{d}s^{2}=-\mathrm{d}\hat{t}^{\,2}+\mathrm{d}\hat{r}^{2}+\left(1-\frac{3X}{8}+ \mathcal{O}(X^2)\right)\hat{r}^{2}\left(\mathrm{d}\theta^2 +\sin^{2}{\theta}\mathrm{d}\phi^2\right).
  \label{reff}
\end{equation}

Observe that this line element is similar to the global monopole one \cite{PhysRevLett.63.341} with a deficit solid angle, $\delta=\frac{3X}{8}$, up to first-order in $X$. As pointed out in the former section, the photon trajectories can be restricted to the equatorial plane (surface $\theta=\pi/2$) due to the spherical symmetry. For this case, the previous line element reduces to
\begin{equation}
     \lim\limits_{\hat{r} \to \infty}\mathrm{d}s^{2}=-\mathrm{d}\hat{t}^{\,2}+\mathrm{d}\hat{r}^{2}+\left(1-\frac{3X}{8}+ \mathcal{O}(X^2)\right)\hat{r}^{2}\mathrm{d}\phi^2 .
     \label{lin}
\end{equation}
Note that this spacetime can be viewed as a cosmic string geometry intersected by $z=const$ planes. As a result, the line element \eqref{lin} is locally flat with conical topology and, of course, it is asymptotically flat. To further clarify this issue, we can define the new angular variable $\hat{\phi}=\phi\left(1-\frac{3X}{8}+\mathcal{O}(X^2)\right)^{1/2}$ and the line element becomes
$$
\lim\limits_{\hat{r} \to \infty}\mathrm{d}s^{2}=-\mathrm{d}\hat{t}^{\,2}+\mathrm{d}\hat{r}^{2}+\hat{r}^{2}\mathrm{d}\hat{\phi}^2,
$$
which is evident flat.

As argued in Refs. \cite{Filho:2022yrk,hassanabadi2023gravitational}, neither the horizon nor the photon sphere are modified due to the LSB process. Therefore, the critical parameter $b_{c}$ is given below
\ie
b_{c} =  3 \sqrt{3}M \frac{\sqrt[4]{((4-X) (3 X+4))}}{2}.
\fe
Also, $\Tilde{a}$ and $\Tilde{b}$ can be expressed as
\ie
\Tilde{a} = \sqrt[4]{\frac{\frac{3 X}{4}+1}{\left(1-\frac{X}{4}\right)^3}},
\fe
and
\ie
\Tilde{b} = \sqrt[4]{\frac{\frac{3 X}{4}+1}{\left(1-\frac{X}{4}\right)^3}} \ln{6} + I_{R}(r_{m}) - \pi.
\fe
In contrast with what happens in the Schwarzschild case, notice that the contribution to the parameter $\Tilde{a}$ is fundamentally due to the feature coming from the LSB. In addition, $I_{R}(r_{m})$ can be calculated as 
\ie
\begin{split}
I_{R}(r_{m}) = & \int_{0}^{1} \sqrt[4]{\frac{\frac{3 X}{4}+1}{\left(1-\frac{X}{4}\right)^3}} \left( \frac{2}{z \sqrt{1 - \frac{2z}{3}}} - \frac{1}{z}\right)   \mathrm{d}z \\
= & 2\sqrt[4]{\frac{\frac{3 X}{4}+1}{\left(1-\frac{X}{4}\right)^3}} \ln [6(2-\sqrt{3})].
\end{split}
\fe
Notice that $\Tilde{b}$ can straightforwardly be written as
\ie
\Tilde{b} = \sqrt[4]{\frac{\frac{3 X}{4}+1}{\left(1-\frac{X}{4}\right)^3}} \ln \left[216 \left(7-4 \sqrt{3}\right)\right] - \pi.
\fe
Therefore, the deflection angle displayed in Eq. (\ref{deflections}) is finally found to have the form
\begin{eqnarray}
a(b) &=& - \sqrt[4]{\frac{\frac{3 X}{4}+1}{\left(1-\frac{X}{4}\right)^3}} \ln \left[   \frac{2b}{3 \sqrt{3}M \sqrt[4]{((4-X) (3 X+4))}} - 1 \right]\nonumber\\ & +& \sqrt[4]{\frac{\frac{3 X}{4}+1}{\left(1-\frac{X}{4}\right)^3}} \ln \left[216 \left(7-4 \sqrt{3}\right)\right] - \pi \\
&+& \mathcal{O}\left\{ \left(b - 3 \sqrt{3}M \frac{\sqrt[4]{((4-X) (3 X+4))}}{2}\right) \ln \left[ b - 3 \sqrt{3}M \frac{\sqrt[4]{((4-X) (3 X+4))}}{2}\right]\right\}.\nonumber
\end{eqnarray}

\begin{figure}
    \centering
    \includegraphics[scale=0.5]{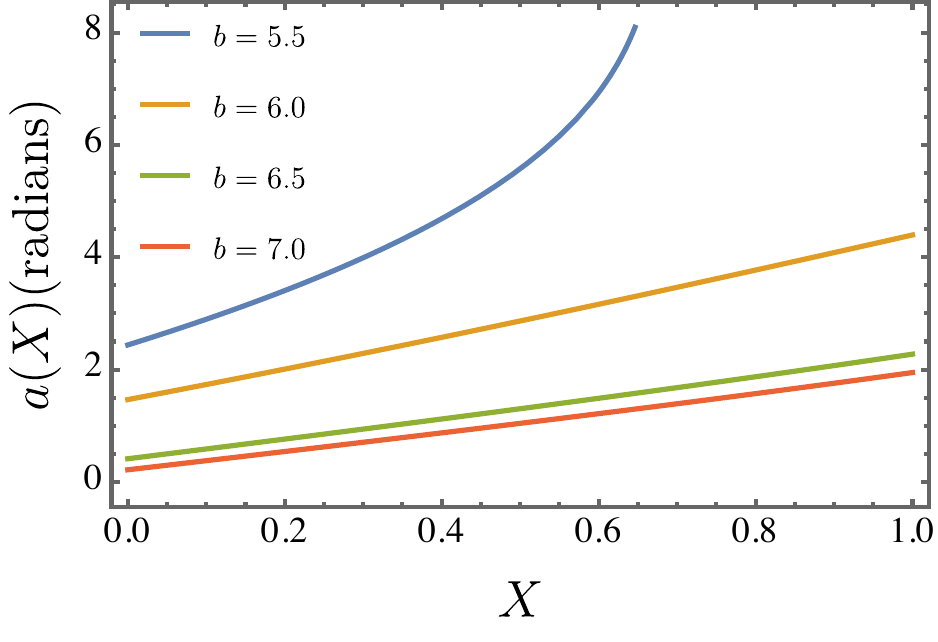}
    \caption{The deflection angle as a function of $X$ for different values of $b$ and $M$.}
    \label{plotsdeflection}
\end{figure}

\begin{figure}
    \centering
     \includegraphics[scale=0.52]{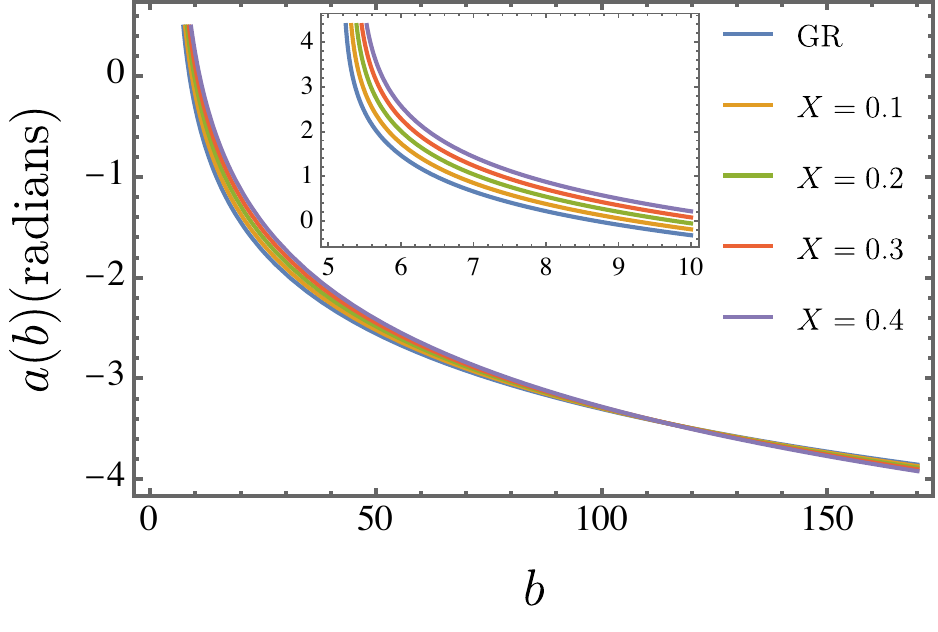}
    \caption{The LV contribution of the deflection $\alpha(b)$ a function of $b$ for different values of $X$. In this case, the GR represents the Schwarzschild solution of the general relativity}
    \label{alphaforX}
\end{figure}

\begin{figure}
    \centering
     \includegraphics[scale=0.62]{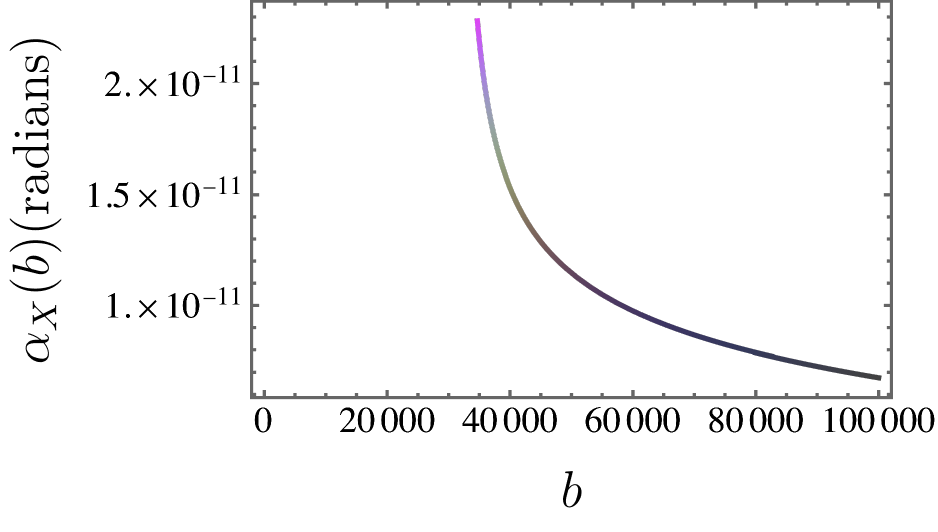}
    \caption{The LV contribution of the deflection $\alpha_{X}(b)$ a function of $b$.}
    \label{alphaX}
\end{figure}

\begin{figure}
    \centering
     \includegraphics[scale=0.52]{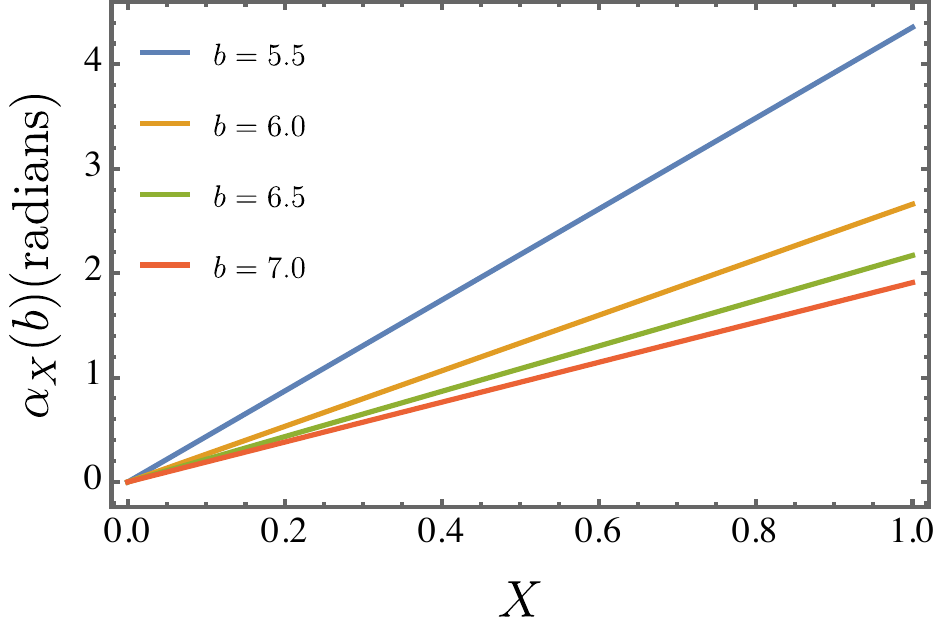}
    \caption{The LV contribution of the deflection $\alpha_{X}(b)$ a function of $X$ for different values of $b$.}
    \label{alphaXfoobb}
\end{figure}

In order to provide a better comprehension to the reader, we display Fig. \ref{plotsdeflection}. Here, we represent the deflection angle as a function of $X$ for different configurations of the system. Furthermore, we present $\alpha(b)$ for various values of $X$, comparing these results with General Relativity (Schwarzschild case) in Fig. \ref{alphaforX}. Also, if we take into account the expansion $a(b)$ in terms of parameter $X$, we obtain
\ie
a_{X << 0}(b) = \alpha_{G}(b) + \alpha_{X}(b),
\fe
where $\alpha_{G}(b)$ and $ \alpha_{X}(b)$ are the contributions due to the usual GR and the LV term $X$, respectively. These terms are given below as follows:
\ie
\alpha_{G}(b) = \ln \left(216 \left(7-4 \sqrt{3}\right)\right) - \ln \left(\frac{b}{3 \sqrt{3} M } -1\right) - \pi,
\fe
and
\ie
\alpha_{X}(b) = \frac{X}{8}  \left(\frac{b}{b-3 \sqrt{3} M}-3 \ln \left(\frac{b}{3 \sqrt{3} M}-1\right)+3 \ln \left(216 \left(7-4 \sqrt{3}\right)\right)\right).
\fe
In Fig. \ref{alphaX}, based on the estimation for the upper bounds of parameter $X$ encountered in Ref. \cite{Filho:2022yrk}, i.e., $X=7.4 \times {10^{-12}}$, and in values of the mass expressed as $4.4 \times 10^6 M_{\odot}$ \cite{genzel2010galactic}, we display the LV contribution for the deflection, $\alpha_{X}(b)$, a function of the impact parameter $b$. Notice that, as $b$ increases, $\alpha_{X}(b)$ decreases. It is worth mentioning that Ref. \cite{Lambiase:2023zeo} recently addressed the gravitational lensing of such a black hole, considering the weak field approximation instead. In addition, we present Fig. \ref{alphaXfoobb}, which highlights the LV contribution to the deflection $\alpha_{X}(b)$ as a function of $X$ for various values of $b$.

\section{Lenses and observables}

\begin{figure}
    \centering
     \includegraphics[scale=0.9]{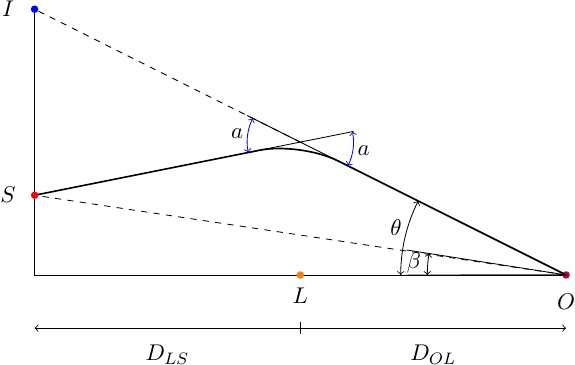}
     \caption{The emitted light from the source, denoted as $S$ (red point), undergoes deflection on its path toward the observer, identified as $O$ (purple point), influenced by the presence of the compact object located at point $L$ (orange point); $I$ (blue point) represents the image seen from the observer $O$; $D_{OL}$ signifies the distance between the lens $L$ and the observer $O$, while $D_{LS}$ denotes the distance between the projection of the source in relation to the optical axis and the lens. }
       	\label{asLrG}
\end{figure}

In this section, we shall explore various parameters associated with the bending of light within the strong gravitational field limit of the LV black hole \eqref{metric3}. To begin with, let us consider Fig. \ref{asLrG} to illustrate a gravitational lensing phenomenon by the black hole \eqref{metric3}. The emitted light from the source, denoted as $S$ (red point), undergoes deflection on its path toward the observer, identified as $O$ (purple point), influenced by the presence of the LV black hole located at point $L$ (orange point). It is worth mentioning that $I$ (blue point) represents the image seen from the observer $O$. The angular positions of the source and the observed image are designated as $\beta$ and $\theta$, respectively. The angular deviation of light, denoted by $a$, encapsulates the alteration in the light's trajectory as it traverses through this gravitational field.

In addition, we employ the identical arrangement as proposed in \cite{030,bozza2001strong}, wherein we assume that the source ($S$) exhibits near--perfect alignment with the lens ($L$). This specific scenario is noteworthy for the presence of relativistic images. Under these conditions, the lens equation governing the relationship between $\theta$ and $\beta$ is expressed as follows:
\ie
\label{EqLente}
\beta=\theta-\frac{D_{LS}}{D_{OS}}\Delta a_{n}.
\fe
Here, $\Delta a_{n}$ represents the deflection angle after accounting for all the loops completed by the photons before reaching the observer, specifically given by $\Delta a_{n}=a-2n\pi$. In this methodology, we adopt the following approximation for the impact parameter: $\Tilde{b}\simeq\theta D_{OL}$. Consequently, we can express the angular deviation as follows:
\ie
\label{defle}
a(\theta)=- \Tilde{a}\ln\left(\frac{\theta D_{OL}}{b_c}-1\right)+\Tilde{b}.
\fe

In order to derive $\Delta a_{n}$, we undertake an expansion of $a(\theta)$ around $\theta=\theta^{0}_n$, where the condition $\alpha(\theta^{0}_n)=2n\pi$ holds:
\ie
\label{da}
\Delta a_{n}=\frac{\partial a}{\partial\theta}\Bigg|_{\theta=\theta^0_n}(\theta-\theta^0_n) \ .
\fe
Taking into account Eq. (\ref{defle}) at $\theta=\theta^{0}_n$, we obtain:
\ie
\label{To}
\theta^0_{n}=\frac{b_c}{D_{OL}}\left(1+e_n\right), \qquad\text{where}\quad e_n=e^{\Tilde{b}-2n\pi} \ .
\fe

By substituting (\ref{To}) into (\ref{da}), we obtain $\Delta a_{n}=-\frac{\Tilde{a}D_{OL}}{b_ce_n}(\theta-\theta^0_n)$. Further incorporating this result into the lens equation (\ref{EqLente}), we derive the expression for the $n^{th}$ angular position of the image
\ie
\theta_n\simeq\theta^0_n+\frac{b_ce_n}{\Tilde{a}}\frac{D_{OS}}{D_{OL}D_{LS}}(\beta-\theta^0_n) \ .
\fe

While the deflection of light maintains surface brightness, the gravitational lens introduces changes to the solid angle of the source, influencing its appearance. The total flux received from a relativistic image is proportionate to the magnification $\mu_{n}$, defined as $	\mu_n=\left|\frac{\beta}{\theta}\frac{\partial\beta}{\partial\theta}|_{\theta^0_{n}}\right|^{-1}$. Utilizing (\ref{EqLente}) and recalling that $\Delta a_{n}=-\frac{\bar{a}D_{OL}}{b_ce_n}(\theta-\theta^0_n)$, we obtain:
\ie
\mu_{n}=\frac{e_n(1+e_n)}{\Tilde{a}\beta}\frac{D_{OS}}{D_{LS}}\left(\frac{b_c}{D_{OL}}\right)^2 \ .
\fe

It should be observed that the magnification factor $\mu_n$ increases as $n$ grows. Consequently, the luminosity emanating from the initial image $\theta_1$ significantly overshadows that of subsequent images. Notably, the overall luminosity remains subdued, primarily due to the presence of the term $\left(\frac{b_c}{D_{OL}}\right)^2.$ A noteworthy observation is the emergence of magnification divergence as $\beta \to 0$, emphasizing that the optimal alignment between the source and the lens maximizes the potential for detecting relativistic images.

In summary, we have explicated the positions and fluxes of relativistic images in terms of expansion coefficients ($\Tilde{a}$, $\Tilde{b}$, and $b_c$). Shifting our focus to the inverse problem, our aim is to discern these expansion coefficients from empirical observations. This pursuit not only facilitates a comprehensive understanding of the nature of the object responsible for generating the gravitational lens but also enables nuanced comparisons with predictions derived from modified theories of gravity.

In addition, the impact parameter can be expressed in relation to $\theta_{\infty}$, as detailed in \cite{bozza2001strong}
\ie
b_c=D_{OL}\theta_{\infty} \ ,
\fe
where $\theta_{\infty}$ represents the other relativistic images.
We shall follow Bozza's methodology, as outlined in \cite{bozza2001strong}, where the resolution considers only the outermost image $\theta_{1}$ as a distinct entity, with the remaining images encompassed within $\theta_{\infty}$. To give more insight on this, Bozza introduced the following observables:
  \begin{eqnarray}
       s&=&\theta_{1}-\theta_{\infty}= \theta_{\infty} e^{\frac{\Tilde{b}-2\pi}{\bar{a}}} \ ,\\
       \tilde{r}&=& \frac{\mu_{1}}{\sum\limits_{n=2}^{\infty} \mu_{n} }= e^{\frac{2\pi}{\Tilde{a}}} \ .
       \end{eqnarray}
In the aforementioned expressions, $s$ denotes the angular separation, and $\tilde{r}$ represents the ratio of the flux from the first image to the combined flux of all other images. These formulations can be inverted to derive the expansion coefficients. In the subsequent subsection, we will present a specific astrophysical example to compute the mentioned observables and evaluate the influence of the LV parameter, $X$, on these quantities.

\subsection{Galactic phenomena: gravitational lensing by Sagittarius A$^{*}$}

Analysis of observational data on stellar dynamics reveals the compelling presence of a dense, enigmatic entity nestled within the heart of our galaxy. This enigmatic entity is believed to be a supermassive black hole, called Sagittarius (Sgr) A$^{*}$, with its mass estimated to be a staggering $4.4 \times 10^6 M_{\odot}$ \cite{genzel2010galactic}. In our quest to understand celestial phenomena, we delve into the characteristics of this astronomical phenomenon, leveraging the dimensionless parameter $X$ to elucidate the behavior of observables.

\begin{figure}
    \centering
     \includegraphics[scale=0.41]{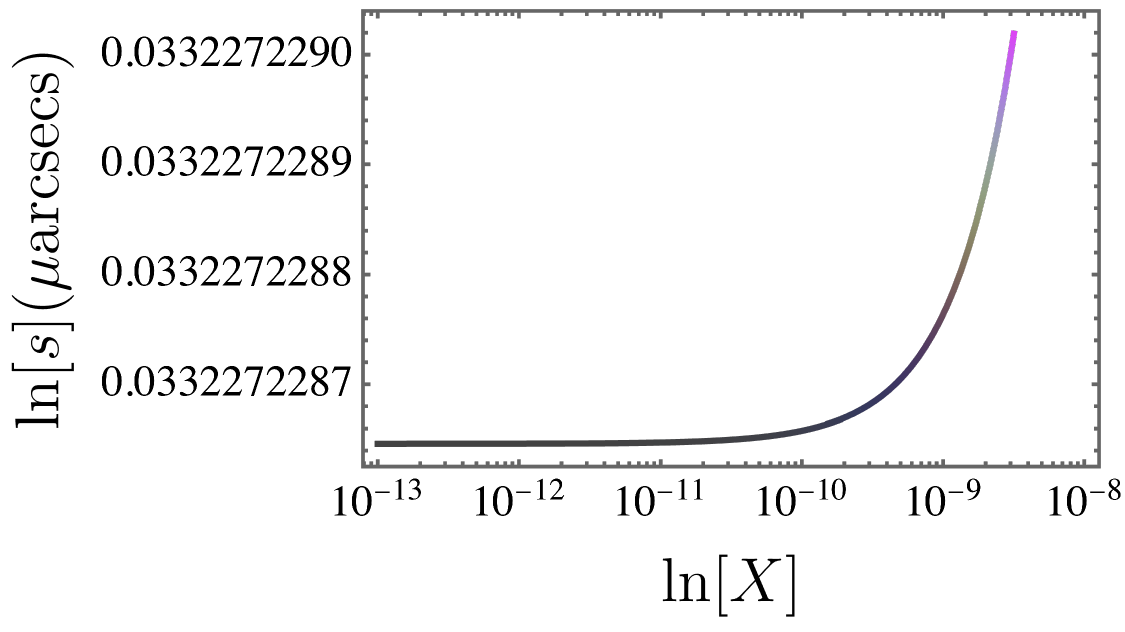}
    \caption{Observable $\ln s$ for different values of $\ln X$.}
    \label{sandrobsevables}
\end{figure}

\begin{figure}
    \centering
     \includegraphics[scale=0.5]{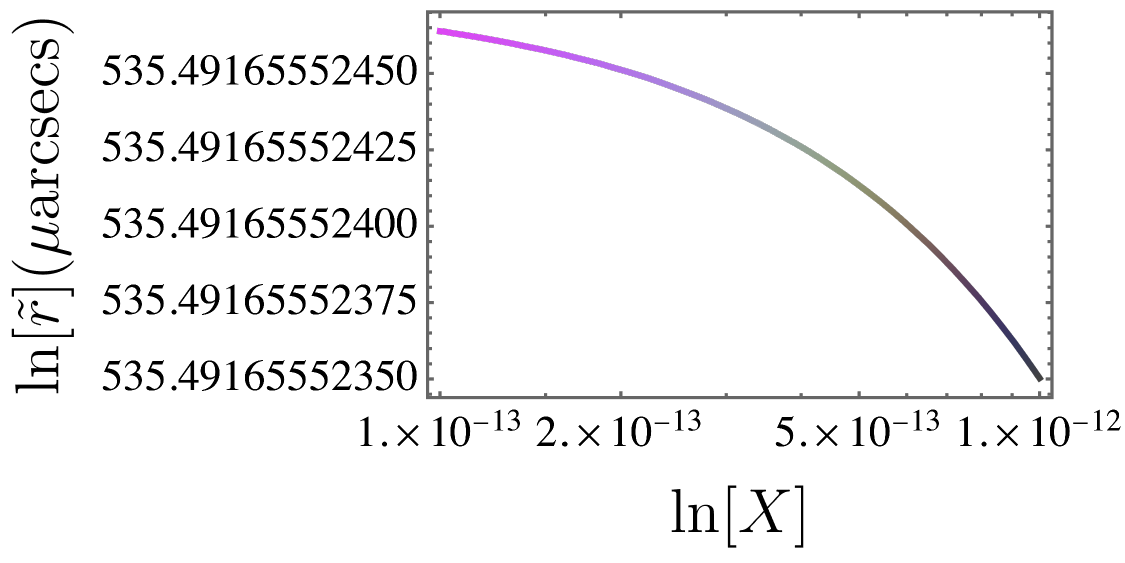}
    \caption{Observable $\ln r$ for different values of $\ln X$.}
    \label{sandrobsevables2}
\end{figure}

To evaluate the observables, let us take distance $D_{OL} = 8. 5 \text{Kpc}$ and $X \sim 7.4 \times 10^{-12}$, accordingly the literature \cite{genzel2010galactic,Filho:2022yrk}. In possessing with $b_{c} = 3 \sqrt{3}M \sqrt[4]{(4-X) (3 X+4)}/2$, we can obtain $\theta_{\infty} \approx 26.55\mu \text{arcsecs}+ \mathcal{O}(X)$, where $\mathcal{O}(X)$ indicates the first-order in the LV parameter which is of the order of $10^{-12}$. 
In order to provide a better comprehension for the reader, we display Figs. \ref{sandrobsevables} and \ref{sandrobsevables2}, representing the two observables, i.e., $s$ and $\Tilde{r}$ in terms of the LV coefficient. Note that we have used logarithmic scales due to fact that the LV contribution is very small. A direct inspection of the plot \eqref{sandrobsevables} shows that the observable $s$ increases as $X$ grows which entails that the first image is further away from the other relativistic images. On the other hand, the plot \eqref{sandrobsevables2} displays a different behavior, i.e., the ratio of the flux of the first image decreases as $X$ grows.

\section{Summary and conclusion}
\label{summary}

Being motivated by our previous paper on the LV black hole \cite{Filho:2022yrk}, our study focused on a specific task, namely, the computation of gravitational lensing effects under the strong field limit. Notably, our analytical derivation of the deflection angle enabled \textit{analytical} result of the lens equation. This methodological rigor not only bolstered the accuracy of our computations but also yielded tangible outcomes. Consequently, we were able to ascertain relativistic image positions and their associated magnifications with heightened precision, thereby contributing significantly to the broader understanding of gravitational lensing phenomena.

Within this paper,  we explored the intricacies of relativistic image properties and employed expansion coefficients, i.e., $\Tilde{a}$, $\Tilde{b}$, and $b_c$, to evaluate their positions and fluxes. Transitioning to the inverse problem, our primary objective was to extract these coefficients from empirical observations. This pursuit not only deepened our understanding of gravitational lensing mechanisms but also facilitated nuanced comparisons with predictions derived from modified gravitational theories. Furthermore, we performed computations for observables $s$ and $\Tilde{r}$, unveiling their transformations in terms of the dimensionless LV parameter $X$.

Furthermore, we provided an application of our results, concentrating on the galactic phenomena of a supermassive black hole, taking Sagittarius A$^{*}$ as a natural example. In this case, some estimations for $b_{c}$ and $\theta_{\infty}$ were accomplished as well based on the upper bounds of $X$ encountered in \cite{Filho:2022yrk}. In addition, we provided the behavior of the observables $s$ and $\tilde{r}$ in terms of $X$ and we found that $s$ grows as $X$ grows while $\tilde{r}$ decreases as $X$ grows.

The natural continuation of our study could consist in applying this methodology to other black hole solutions arising within various LV extensions of gravity, in particular regular and rotating \cite{Filho:2024hri} ones. We plan to perform these studies in forthcoming papers.

\section*{APPENDIX}

In this Appendix, we present some basic definitions of the Lorentz symmetry breaking. 

Let us discuss the distinction between \emph{observer} and \emph{particle} Lorentz transformations in the Standard Model Extension (SME) framework developed by Kostelecký et al.,  see \cite {Colladay:1996iz,Colladay:1998fq}. Since our paper deals with gravitational lensing in a bumblebee model formulated in the metric–affine approach, we shall provide concise explanations and citations to guide readers unfamiliar with these preliminaries.  The essential ideas are summarized below.

\subsection*{1.\;Observer versus particle Lorentz transformations}

A fixed background vector $b^{a}$ (or, in general, any fixed background tensor) selects a preferred direction in each local Lorentz frame.  
Under an \emph{active} boost or rotation the dynamical fields transform as usual, while $b^{a}$ does not; the action therefore changes {\bf signalizing} broken particle Lorentz symmetry.  
Under a \emph{passive} coordinate change, all components transform together, so the action keeps its form, and the observer Lorentz symmetry remains intact. 

\subsection*{2.\;Role of the coefficient $X$ in the solution}

In the bumblebee model, the vector field acquires the vacuum value $\langle B_{\mu}\rangle = b_{\mu}$, enforced by a potential of the form $V(B_{\mu}B^{\mu}\mp b^{2})$.  
Solving the modified Einstein equations {\bf gives} the metric displayed in Eq.\,(\ref{metric3}).  
All Lorentz-violating contributions enter through the dimensionless combination
\[
X \;=\; \xi\, b^{2},
\]
which rescales the temporal and angular metric functions. Once $b^{\mu}$ is radial and static, spherical symmetry is preserved, and lensing analysis proceeds exactly as in the Schwarzschild case after the redefinitions in Eqs.\eqref{14}, \eqref{15} and \eqref{metric4}.

\medskip
\noindent
\textbf{\underline{Observer Lorentz invariance}:}  Horizon radius, photon-sphere radius, deflection angle, and other observables are scalars under passive coordinate changes; they depend on $X$ only through the metric and geodesic equations, which transform covariantly.

\medskip
\noindent
\textbf{\underline{Broken particle Lorentz invariance}:}  An active boost in the local frame changes the tetrad components of $b^{a}$ while leaving $b^{2}$ unchanged.  
The field equations must then be solved again with the new orientation, and the resulting deflection angle $\alpha$ generally differs from the one derived in Sec.\,III.  

This framework is used within our paper.

\section*{Data Availability Statement}

This manuscript has no associated data.

\section*{Acknowledgments}
\hspace{0.5cm}The authors would like to thank the Conselho Nacional de Desenvolvimento Cient\'{\i}fico e Tecnol\'{o}gico (CNPq) for financial support. P. J. Porf\'{\i}rio would like to acknowledge the Brazilian agency CNPq, grant No. 307628/2022-1. The work by A. Yu. Petrov has been partially supported by the CNPq project No. 303777/2023-0. Moreover, A. A. Araújo Filho is supported by Conselho Nacional de Desenvolvimento Cient\'{\i}fico e Tecnol\'{o}gico (CNPq) and Fundação de Apoio à Pesquisa do Estado da Paraíba (FAPESQ), project No. 150891/2023-7.

\bibliographystyle{ieeetr}
\bibliography{main}

\end{document}